\documentclass[twocolumn]{emulateapj}

\usepackage{graphicx}
\usepackage{color}
\usepackage[colorlinks=true,linkcolor=blue,citecolor=blue]{hyperref}

\begin{document}

\title{Survival of the fittest: Numerical modeling of
Supernova 2014c}

\shorttitle{Numerical modeling of
Supernova 2014c}

\author{
Felipe Vargas\altaffilmark{1,2}, 
Fabio De Colle\altaffilmark{1}, 
Daniel Brethauer\altaffilmark{3},
Raffaella Margutti\altaffilmark{3}
\& Cristian G. Bernal\altaffilmark{2} 
}
\altaffiltext{1}{Instituto de Ciencias Nucleares, Universidad Nacional Aut\'onoma de M\'exico, A. P. 70-543 04510 D. F. Mexico}

\altaffiltext{2}{Universidade Federal do Rio Grande, Av. Italia km 8 Bairro Carreiros, Rio Grande, RS, Brazil}

\altaffiltext{3}{Center for Interdisciplinary Exploration and Research in Astrophysics (CIERA) and Department of Physics and Astronomy, Northwestern University, Evanston, IL 60208}


\begin{abstract}
Initially classified as a supernova (SN) type Ib, $\sim$ 100 days after the explosion SN\,2014C made a transition to a SN type II, presenting a gradual increase in the H${\alpha}$ emission. This has been interpreted as evidence of  interaction between the supernova shock wave and a massive shell previously ejected from the progenitor star. In this paper, we present numerical simulations of the propagation of the SN shock through the progenitor star and its wind, as well as the interaction of the SN ejecta with the massive shell. To 
determine with high precision the structure and location of the shell, we couple a genetic algorithm to a hydrodynamic and a bremsstrahlung radiation transfer code. We  iteratively modify the density stratification and location of the shell by minimizing the variance between X-ray observations and synthetic predictions computed from the numerical model. By assuming spherical symmetry, we found that the shell has a mass of 2.6 M$_\odot$, extends from 1.6 $\times 10^{16}$ cm to $1.87  \times 10^{17}$ cm, implying that it was ejected $\sim 60/(v_w/100 {\rm \; km \; s^{-1}})$ yrs before the SN explosion, and has a density stratification decaying as $\sim r^{-3}$. We found that the product of metallicity by the ionization fraction (due to photo-ionization by the post-shock X-ray emission) 
is $\sim$ 0.5. Finally, we predict that, if the density stratification follows the same power-law behaviour, the SN will break out from the shell by mid 2022, i.e. 8.5 years after explosion.

\end{abstract}



\section{Introduction}

While mass loss is one of the key mechanisms regulating the evolution of massive stars, a complete understanding of it is still missing, specially during the final phases before the supernova (SN) explosion  \citep[e.g.,][]{smith14}. The mass loss history $\lesssim 100-1000$ yrs before core-collapse supernova explosions can be inferred from the radio and X-ray emission resulting from the propagation of the SN shock through the circumstellar material (for a review see, e.g., \citealt{chevalier17}).
While the bulk of the SN ejecta emits in optical, the shock heated gas resulting from the interaction with the environment might be observed in X-ray and radio due to bremsstrahlung and synchrotron radiation from relativistic electrons accelerated at the shock front.

The forward shock of type Ib/c SNe, originated from the collapse of Wolf-Rayet progenitor stars, interacts with the wind of the progenitor star, which has typical mass loss rates of $\dot{M}_w \sim 10^{-4}-10^{-6}$ M$_\odot$ yr$^{-1}$. The wind is often inhomogenous, as proven by radio emission, showing small flux fluctuations of $\sim$ a few over timescales of tens-hundreds of days after the explosion \citep[e.g.,][]{bietenholz05, soderberg06, schinzel09, wellons12, salas13, bietenholz14, corsi14, palliyaguru19}.

In a few cases, type Ib/c SNe show signs of much stronger interaction between the ejecta and shells of material ejected before the explosion. For instance, SN 2006jc exploded inside a dense He rich environment \citep[e.g.,][]{foley07}, likely produced by an outburst ejected $\sim$ two years before the SN.
SN 2001em, initially classified as type Ib SN, presented prominent H$\alpha$ emission lines at 2.5 yrs. Associated with strong radio and X-ray emission, this was interpreted as evidence of the interaction between the SN ejecta and a massive ($\sim 3$ M$_\odot$) hydrogen-rich shell located at $\sim 7\times 10^{16}$ cm \citep{chugai06, chandra20}. Several other SNe show similar signs of early interaction with massive shells \citep[see, e.g.,][]{anupama05, moriya14, chen18, pooley19, dwarkadas10, Ben-Ami14, Mauerhan18, suzuki21}. 
Nevertheless, while in all these cases the intermediate phases of the transition between type Ib and type IIn SN were not observed, this transition has been observed in detail in the SN\,2014C.

Discovered by the \textit{Lick Observatory Supernova Search} \citep{kim14} in the NGC 7331 galaxy at a distance of 14.7 Mpc and initially classified as a type Ib SN, SN\,2014C made a transition to a type IIn SN  about $\sim 100$ days after the explosion, showing strong H$\alpha$ emission \citep{milisavljevic15}.
The modelling of the optical/UV light curve shows that SN\,2014C has a kinetic energy of 1.75 $\pm$ 0.25 $\times 10^{51}$ erg, with an ejecta mass of 1.7 $\pm$ 0.2 M$_\sun$ and a Nickel mass of 0.15 $\pm$ 0.02  M$_\sun$ \citep{margutti17}.

SN\,2014C has also been extensivel observed with X-rays Telescope (XRT, \citep{Burrows05}) on board the \textit{Neil Gehrels Swift Observatory (XRT)} \citep{Gehrels04} and the \textit{Chandra X-ray Observatory (CXO)}  in the 0.3-10 keV energy band, and by the \textit{Nuclear Spectroscopic Telescope Array (NuSTAR)} from 3 keV to 79 keV \citep{margutti17,brethauer20,brethauer21}. Most of the detected X-ray emission is concentrated in the 1-40 keV energy band, while emission below $\sim 1$ keV is strongly absorbed.

The observed X-ray emission increased at 250 days, although the lack of temporal coverage between 100 and 250 days implies that likely the onset of the X-ray increase was at a smaller time.
Integrated over the spectral range 0.3-100 keV, it raised from $5 \times 10^{39}$ erg s$^{-1}$ to $5 \times 10^{40}$ erg s$^{-1}$.
Then, it peaked $\sim 850 - 1000$ days and maintained a nearly constant flux until $\sim 2000$ days after the explosion \citep{brethauer20}. 

Radio observations showed a similar behaviour. Light curves of the SN at 15.7 GHz taken with the \emph{Arcminute Microkelvin Imager} between 16 and 567 days showed that the flux increased rapidly at $\sim$ 100-150 days \citep{anderson17}. A similar increase (by more than one order of magnitude) was observed by the \emph{Very Large array} \citep{margutti17}.
Furthermore, observations done by using the \emph{Very Long Baseline Interferometry} found that the shock expansion has already strongly decelerated 384 days after the explosion \citep{bietenholz18,bietenholz20}.

Altogether, the evolution of optical, radio and X-ray emission have been interpreted by considering the interaction of the SN ejecta with a wind with $\dot{M}= 5 \times 10^{-5}$ M$_\sun$ yr$^{-1}$ around the progenitor star, and a massive shell located at R$_{\rm sh} = 5 \times 10^{16}$ cm, with an extension of approximately 0.25 R$_{\rm sh}$ and a density of $\sim$10$^6$ cm$^{-3}$ \citep{milisavljevic15, anderson17, margutti17, bietenholz18, bietenholz20}. Assuming a velocity of 10-100 km s$^{-1}$ for the shell, its position implies that it was ejected $\sim 100$-$1000$ yrs before the SN explosion \citep{milisavljevic15, anderson17, margutti17, bietenholz18, bietenholz20}.
Late infrared, optical and UV observations confirm that the SN shock is still interacting with the massive shells five years after the explosion \citep{tinyanont19, son20}.

\begin{figure}
\centering
\includegraphics[height=1.3\linewidth]{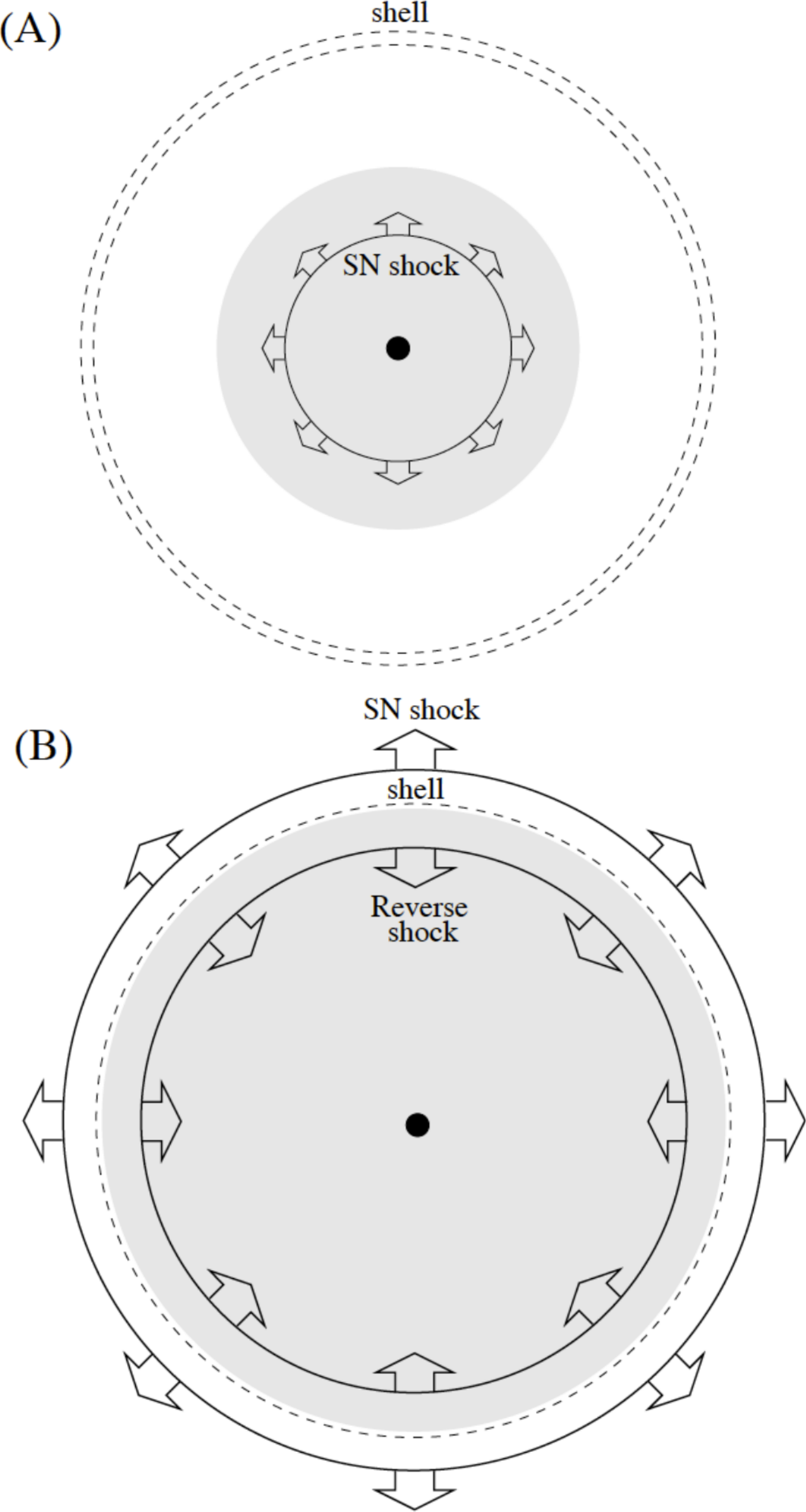}
\caption{Schematic evolution of a supernova
shock interacting with an external shell
(not-to-scale). (A) the supernova explosion
produces a outwardly
propagating shock; (B) the reverse shock becomes stronger as the shock front collides with the external shell, moving back towards (in mass coordinates) the
supernova center.}
\label{fig1}
\end{figure}

Radio and X-ray emission from the SN shock are typically described by considering a self-similar behaviour for the dynamics \citep[e.g.,][]{chevalier82, chevalier89, chevalier98}. In the case of a supernova interacting with a massive shell, we can not assume a self-similar behaviour for the flow as the interaction with the shell leads to the formation of a strong reverse shock (see Figure \ref{fig1}). Then, although an analytical description gives a qualitative picture of the problem, a detailed understanding of the dynamics can be achieved only employing numerical simulations.

Typically, running a large number of models covering a predefined number of parameters provides sufficient detail of the dynamics. When running computational intensive hydrodynamical simulations, however, this approach necessarily limits the parameter space explored. To handle this limitation, in this paper we use an optimization method (a genetic algorithm) to iteratively determine a relatively large number of parameters by running a limited number of numerical simulations. Although there have been a wide range of applications of genetic algorithms in astrophysics \citep[e.g.,][]{Charbonneau1995}, it is the first time that this method has been applied by coupling hydrodynamic simulations with radiative transfer, in order to obtain an optimized solution for the CSM density profiles. 

Specifically, we run a set of numerical simulations coupled to a genetic algorithm (GA). The GA changes the shell density with time by minimizing the difference between the synthetic X-ray emission (computed by post-processing the results of the hydrodynamical simulations) and observations of this SN presented in \citet{margutti17, brethauer20, brethauer21}. In this approach, each density $\rho(r_i)$ (being $i=1$, \dots, $N$) is considered a parameter of the model.

The paper is organized as follows: in Section 2 we describe the hydrodynamics code and the initial conditions of the simulations, the bremsstrahlung radiation transfer code and the genetic algorithm employed to solve the optimization problem and to find the density stratification of the shell. In Section 3 we present the results of our numerical calculations. In Section 4 we discuss the limits of the simulations presented, and the implications of our findings. Finally, in Section 5 we draw our conclusions.

\section{Methods}

\subsection{Numerical codes and initial conditions}

We study the interaction of a SN shock with a massive shell by running a set of one-dimensional (1D), spherically symmetric simulations with the adaptive mesh refinement (AMR) code \emph{Mezcal} \citep{decolle12}. The code solves the special relativistic hydrodynamics equations, and has been extensively used to run numerical simulations of astrophysical flows
\citep[e.g.,][]{gonzalez14, decolle14, decolle18a, decolle18b}.

We follow the propagation of the SN shock front as it moves through a computational grid covering $\sim$ nine orders of magnitude in space, from  $\sim 2\times 10^8$ cm (the outer edge of the iron core) to $\sim 2 \times 10^{17}$ cm. We do so by running two sets of simulations. First, we follow the propagation of the SN shock front as it moves through the progenitor star, e.g., from $\sim 10^8$ cm to $10^{11}$ cm. Then, after remapping the results of the small scale simulation into a much larger computational box, we follow its propagation through the wind of the progenitor star and its interaction with the massive shell located at $\gtrsim 10^{16}$ cm.

In the small scale simulation, we set the density profile of the progenitor star by using the E25 pre-supernova model from \citet{heger20}. This corresponds to a star 
which has lost its hydrogen and helium envelope. The resulting Wolf-Rayet star has a mass of 5.25 M$_{\odot}$ and a radius of $3 \times 10^{10}$ cm. 
The computational grid extends radially from $2.2 \times 10^8$ cm to  $6.6 \times 10^{11}$ cm. We employ 20 cells at the coarsest level of refinement, with 22 levels of refinement, corresponding to a resolution of $1.6\times 10^4$ cm. The SN energy ($E_{\rm SN} \approx 10^{51}$ erg) is imposed by setting the pressure of the two inner cells of the computational box as $p=E_{\rm SN}(\Gamma_{\rm ad}-1)/V$, being $\Gamma_{\rm ad}=4/3$ the adiabatic index and $V$ the volume of the two cells. Outside the stellar surface, we take $\rho=\dot{M}_w/(4\pi r^2 v_w)$, being
$\dot{M}_w = 5 \times 10^{-6}$ M$_\odot$ yr$^{-1}$ and $v_w=10^8$ cm s$^{-1}$ the mass loss rate and velocity of the wind launched by the Wolf-Rayet star before the collapse. As the velocity of the SN shock front is about two orders of magnitude larger than the Wolf-Rayet wind (i.e., $\sim 10^{10}$ cm s$^{-1}$ vs. $\sim 10^{8}$ cm s$^{-1}$), we assume that the wind medium is static. The propagation of the SN shock front is followed as it breaks out of the progenitor star and arrives to $4.5 \times 10^{11}$ cm in 50 seconds. 

In the large scale simulations, the computational box goes from $10^{10}$ cm to $5 \times 10^{17}$ cm. For $r < 4.5 \times 10^{11}$ cm we set the density, pressure and velocity by using the values determined from the small scale simulation. For larger radii, we take the density stratification as $\rho=\dot{M}_w/(4\pi r^2 v_w)$, with $\dot{M}_w=5 \times 10^{-6} $ and $v_w=10^8$ cm s$^{-1}$ as in the small scale simulation. We employ 150 cells, with 20 levels of refinement in the AMR grid, corresponding to a resolution of $2.5\times 10^9$ cm. By running different simulations in which we change the number of levels of refinement between 14 and 22 levels, we verify that 20 levels of refinement are enough to achieve convergence.

Trying to reproduce the observed X-ray emission, we have first considered a massive, uniform cold shell located at the radius $R_s$, with mass $M_s$ and thickness $\Delta R_s$.
To determine the best values for these three parameters, we run a grid of $1815$ models by using 11 values of $M_s$ (in the range 1.5 $M_{\odot}$ to 2.5 $M_{\odot}$), 15 values of $\Delta R_s$ (ranging from $7.5 \times 10^{14}$ up to $2.5 \times 10^{15}$ cm), and 11 values of $R_s$ (from $1.8 \times 10^{16}$ to $7.8 \times 10^{16}$ cm). To compare the results of the numerical simulations with observations, we have then employed a ray-tracing code (see Section \ref{free}).
Unfortunately, none of these models give an acceptable fit to the observations, implying that the density of the massive shell is not constant.

To find the density at several radii is a task that is not possible to achieve with a grid of numerical models. For instance, a grid of ten values for the density at 10 different radii implies running $10^{10}$ simulations. 
Thus, to determine the density stratification of the shell, we decided to solve the ``full'' optimization problem. 
This is done by coupling the \emph{Mezcal} code with other two codes: a radiation transfer code which computes the bremsstrahlung radiation (see Section \ref{free}), and a genetic algorithm (described in Section \ref{GA}) that automatically and randomly changes the density profile inside the shell by minimazing the variance between the synthetic observations computed from the numerical model and the X-ray observations.

\subsection{Bremsstrahlung Emission Code}
\label{free}

We post-process the results of the numerical simulations by computing synthetic spectra.  
The specific flux is given by
\begin{equation}
F_{\nu} = \int I_{\nu} \cos \theta d \Omega\;,
\end{equation}
where $d \Omega =  2 \pi \sin \theta d \theta / D^2 $, being $D = 14.7$ Mpc the luminosity distance from the SN\,2014C \citep{freedman01}, and $I_\nu$ is the specific intensity.

The observed X-rays radiation is due to thermal bremsstrahlung emission caused by the interaction between the SN shock and a massive shell \citep{margutti17}. To determine the specific intensity, we solve the radiation transfer equation
\begin{equation}
  \frac{d I_{\nu}}{d \tau_{\nu}} = S_{\nu} - I_{\nu}\;,
\end{equation}
where $S_{\nu}=j_{\nu}/\alpha_{\nu}$ is the source function, $\tau_\nu = \int \alpha_\nu dl$ is the optical depth, and $j_\nu$ and $\alpha_\nu$ are the emissivity and the absorption coefficient respectively. In addition to the bremsstrahlung self-absorption, we also consider photo-electric absorption, which at early times dominates absorption for frequencies $\lesssim 10^{18}$ Hz, so that
\begin{equation}
\alpha_{\nu} = \alpha_{\nu, {\rm ff}} + \alpha_{\nu, {\rm bf}}\;, \qquad j_{\nu} = j_{\nu, {\rm ff}} \;.
\label{alphaeff}
\end{equation}

To compute the bound-free absorption, we use tabulated cross-sections for solar metallicity  \citep{morrison83}.
For the bremsstrahlung coefficients we take 
\citep[e.g.,][]{rybicki86}
\begin{eqnarray}
j_\nu = \frac{6.8}{4 \pi} \times 10^{-38} Z^2 n_e n_i T^{-1/2} e^{-\frac{h \nu}{kT}} G \\
\alpha_\nu = 3.7 \times 10^{8}T^{-1/2} \frac{Z^2 n_e n_i}{\nu^{3}}  \left(1 - e^{-\frac{h \nu}{kT}}\right) G\;, 
\label{jalpha}
\end{eqnarray}
where $Z$, $n_e$ and $n_i$ are the atomic number, electron and ion densities respectively, and $G$ is the Gaunt Factor ($\sim 1$ for the range of parameters considered here). 
All other variables have their usual meaning and everything is in cgs units.
The electron/ion densities and the temperature are directly determined from the numerical simulations by assuming solar metallicity and composition for an ideal gas.

Cooling is not included in the numerical simulations as the bremsstrahlung cooling timescale is $t_{\rm ff} = 60 (T/10^8 {\rm \,K}) (n_e/10^6 {{\rm cm}^{-3}})$ yrs, so it is much larger than the timescales studied here.

\subsection{Genetic Algorithm}
\label{GA}

\begin{figure}
\centering
\includegraphics[height=0.5\linewidth]{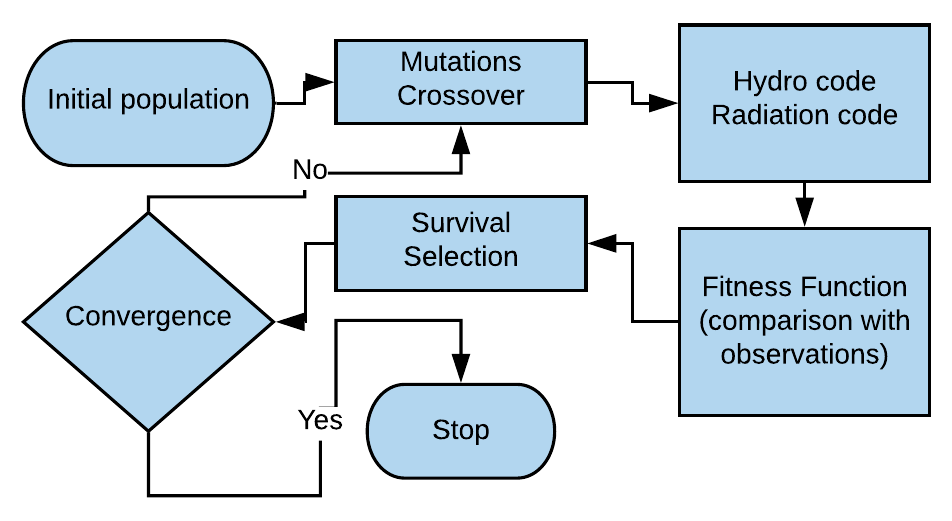}
\caption{Flowchart showing the genetic algorithm employed to determine the density stratification of the massive shell interacting with the SN\,2014C. First, we initialize a population of random densities at different shell radii. We modify the initial population by mutations and cross-over (see the text for details). 
We run hydrodynamical simulations following the interaction of the SN ejecta with the shell and we compute synthetic X-ray spectra which are compared with the observations. We select the ``best'' elements of the population using a ``fitness function'' (the $\chi^2$ test). The process is repeated a thousand times.}
\label{fig2}
\end{figure}

Genetic algorithms \citep[GA hereafter, e.g.,][]{rajpaul12} are based on the theory of natural selection. GA are commonly used optimization methods\footnote{We have employed GA in this paper but the results obtained are independent on the particular  optimization method chosen.}, employed also in astrophysics to solve problems with many degrees of freedom, in which finding the  optimal solution would be very hard otherwise 
\citep[e.g.,][]{canto09, degeyter13, morisset16}.
Nevertheless, this is the first time, as far as we know, that optimization methods are coupled directly to hydrodynamical simulations.

As mentioned before, we made a single small scale simulation for the propagation of the shock-wave in the interior of the star, and we applied the GA to find the best density stratification of the shell by running several simulations of the interaction of the SN ejecta with the massive shell. 

Figure \ref{fig2} shows schematically the GA algorithm implemented. A population formed by 10 elements (the ``chromosomes'' in the GA terminology) is chosen at the beginning of the iterative process. Each element of the population is defined by setting 11 values of the density at different radii inside the shell (the ``genes''). 
Nine of the densities are defined at a radius given by each of the nine observational epochs available. Additionally, we define two densities, one at a smaller and one at a larger radius. At each step, we create 90 new elements of the population. Half of them are defined by randomly choosing two elements of the original population and applying to them cross-over and mutation (see below), while the other half is initialized by directly copying the density values from one random element of the population and modifying it only by mutation. 

In the cross-over process, the two ``parents'' are mixed by choosing randomly a certain number of densities from each element (e.g., the first and second densities from the first element, the third density from the second element and so on). This process is inspired by the genetic mixing present in biological evolution. In the mutation process we modify randomly one density in each element. We do so by setting a Gaussian distribution around the original value of the density $\rho_0$, with a width given, in 90\% of the cases, by $\rho_0/2$, and in 10\% of the cases by $10 \rho_0$, so that in a few cases the system explores density values far away from the initial one (to avoid being trapped by a local minimum).

The shell densities were then mapped as initial conditions into the HD code. Then, after the simulation was completed, the bremsstrahlung X-ray emission was computed by post-processing the results of the HD calculation. 
A fitness function (a reduced $\chi^2$ test) was applied to compare the synthetic spectra produced by the model and the observational data at 9 epochs: 308, 396, 477, 606, 857, 1029, 1257, 1574 and 1971 days after explosion.
The fittest 10 elements (out of the new 90 and the old 10 elements of the population) were saved and used as initial condition for a new step. 
This process was repeated for $\sim 100$ iterations, for a total number of $\sim$ 10$^4$ simulation. Each simulation took $\sim 10$ minutes so that the entire process could be completed in less than a day.

The simulations were done on a cluster of CPUs, by using the ``Message Passing Interface'' library. At each iteration, the master node initialized and synchronised the simulations, selected the best elements and managed the cross-over/mutation processes, while the other nodes run (in parallel) each of the 90 hydrodynamics simulation, compute the X-ray spectra and the fitness function (a $\chi^2$ test).

\section{Results}

\subsection{SN shock propagation through the progenitor wind}
\label{Dy}

\begin{figure}
\centering
\includegraphics[height=1.3\linewidth]{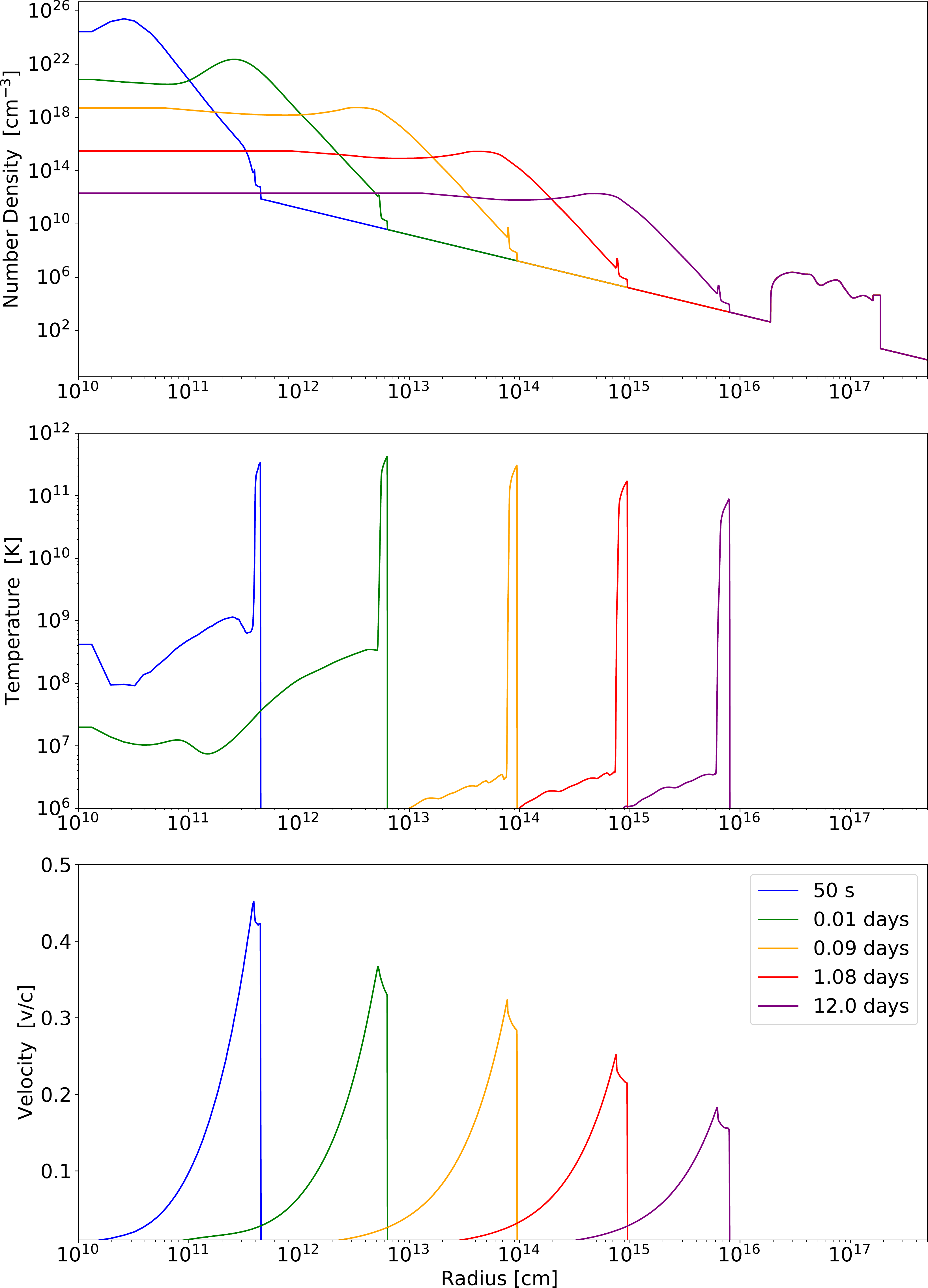}
\caption{Time evolution of the SN shock as it moves through the wind of the progenitor star. From top to bottom: density, temperature and velocity profiles of the SN shock as it interacts with the wind of the progenitor star (at $r \lesssim 1.6\times 10^{16}$ cm) and the outer shell (at $\gtrsim 1.6\times 10^{16}$ cm). }
\label{fig3}
\end{figure}

The propagation of the shock wave through the star has been  extensively studied both for the non-relativistic \citep[e.g.,][]{sakurai60, matzner99} and mildly relativistic regime \citep[e.g.,][]{tan01, Nakar10, Waxman17}.
As the shock approaches the surface of the progenitor star and it moves in the stellar envelope, in which the density drops steeper than $\rho\propto r^{-3}$, the shock velocity increases to mildly relativistic speeds (see the blue line in the bottom panel of Figure \ref{fig3}).

As a result, once the SN shock breaks out in $\sim 25$ s, most of the mass (and energy) of the SN moves at velocities $\sim 10^4$ km s$^{-1}$, while a small fraction of the mass (corresponding to a kinetic energy $\approx 10^{47}$ ergs) expands with larger velocities (up to $v_{\rm sh} \sim 0.5$ c in our simulations).  
Self-similar solutions describing the propagation of the SN shock through a politropic envelope (with $\rho_{\rm env}\propto (R/r-1)^k$) predict an ejecta density stratification $\rho\propto r^{-n}$ after the  break-out, with $n=7-11$, depending on the structure of the progenitor star. As we employ a realistic model for the progenitor star structure (which slightly differs from a politrope), we get $n\sim10$ in the inner part of the ejecta, and a transition to a less steep profile as we approach the shock front (with $n\sim 9$).

Then, we run a set of simulations by using as input the outcome of the small scale simulations, i.e., density, pressure and velocity profiles (blue lines in Figure \ref{fig3}). 
As described in section \ref{GA}, we compute the shell density profile that minimizes the variance between the bremsstrahlung emission computed from the numerical model and the observations.
In the following, we discuss the evolution of the SN shock front while it propagates into the progenitor wind and interacts with the shell density profile obtained after the GA algorithm has converged.

Figure \ref{fig3} shows the evolution of the ejecta before interacting with the outer shell. 
The expansion of the SN ejecta through the progenitor wind leads to the formation of a double shock structure, formed by the forward shock (FS), which accelerates and heats the WR wind, and the reverse shock (RS), which decelerates and heats the SN ejecta. 
Following \citet{chevalier82}, the evolution of the 
forward and reverse shocks is self-similar, with $R \sim t^m$, where $m = (n-3)/(n-2)$. By taking  $n\sim 9$, we get $m\sim 8.5$, which is consistent with the evolution of the shock wave obtained in our simulation (see Figure \ref{fig3}, bottom panel, and Figure \ref{fig4}).
The FS velocity is much larger than the wind velocity. Thus, the post-shock temperature achieves values $\gtrsim 10^{11}$ K, while the SN bulk temperature quickly drops by adiabatic expansion (heating by Ni$^{56}$ and Co$^{56}$ decay is not included in the calculation).

In our simulations, the SN shock acceleration stops only when the shock arrives to the very edge of the progenitor star. This leads to an overestimation of the true shock velocity, as the shock acceleration should stop once the stellar envelope becomes optically thin to the radiation coming from the post-shock region. A detailed calculation of the shock acceleration and break-out is an open problem, which requires a radiation hydrodynamics code.
Once the SN ejecta interacts with the shell (see below), the shock velocity quickly drops. Then, the late evolution of the system will be independent on particular  shock velocity obtained, while it? will depend strongly on the pre-shock structure of the ejecta.

\subsection{SN shock interaction with the massive shell}
\label{SNshell}

\begin{figure}
\centering
\includegraphics[height=0.6\linewidth]{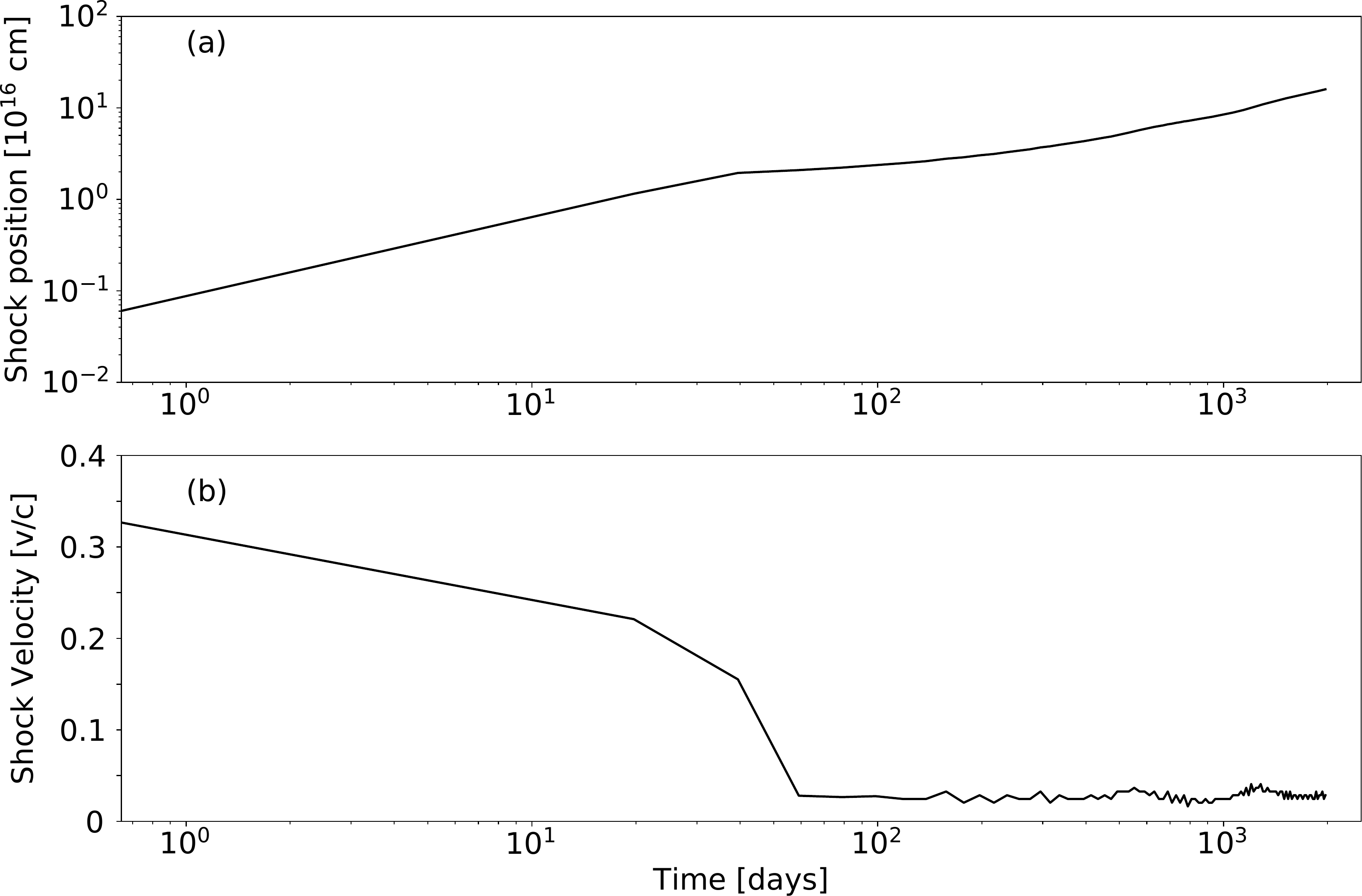}
\caption{\emph{Top panel}: position of the SN shock as function of time. The break seen at $\sim 50$ days corresponds to the beginning of the interaction with the shell. \emph{Bottom panel}: shock velocity.
During the self-similar phase, the shock velocity drops from  $\sim 0.35$c to $\sim 0.15$c. After interacting with the shell, it drops to $\sim 0.025$c.}
\label{fig4}
\end{figure}

At $\sim 50$ days after the explosion, the shell begins interacting with the massive shell (see Figure \ref{fig4}). Radio and optical observations showed that the interaction started $\sim 100$ days after the explosion \citep{milisavljevic15, anderson17}. Then, our simulations overestimate the average shock velocity by a factor of $\sim 2$. A lower shock velocity can be due, as mentioned above, to the loss of thermal energy during the shock breakout or, alternatively, to a less steep density profile in the outer layers of the progenitor stars.

\begin{figure}
\centering
\includegraphics[height=1.3\linewidth]{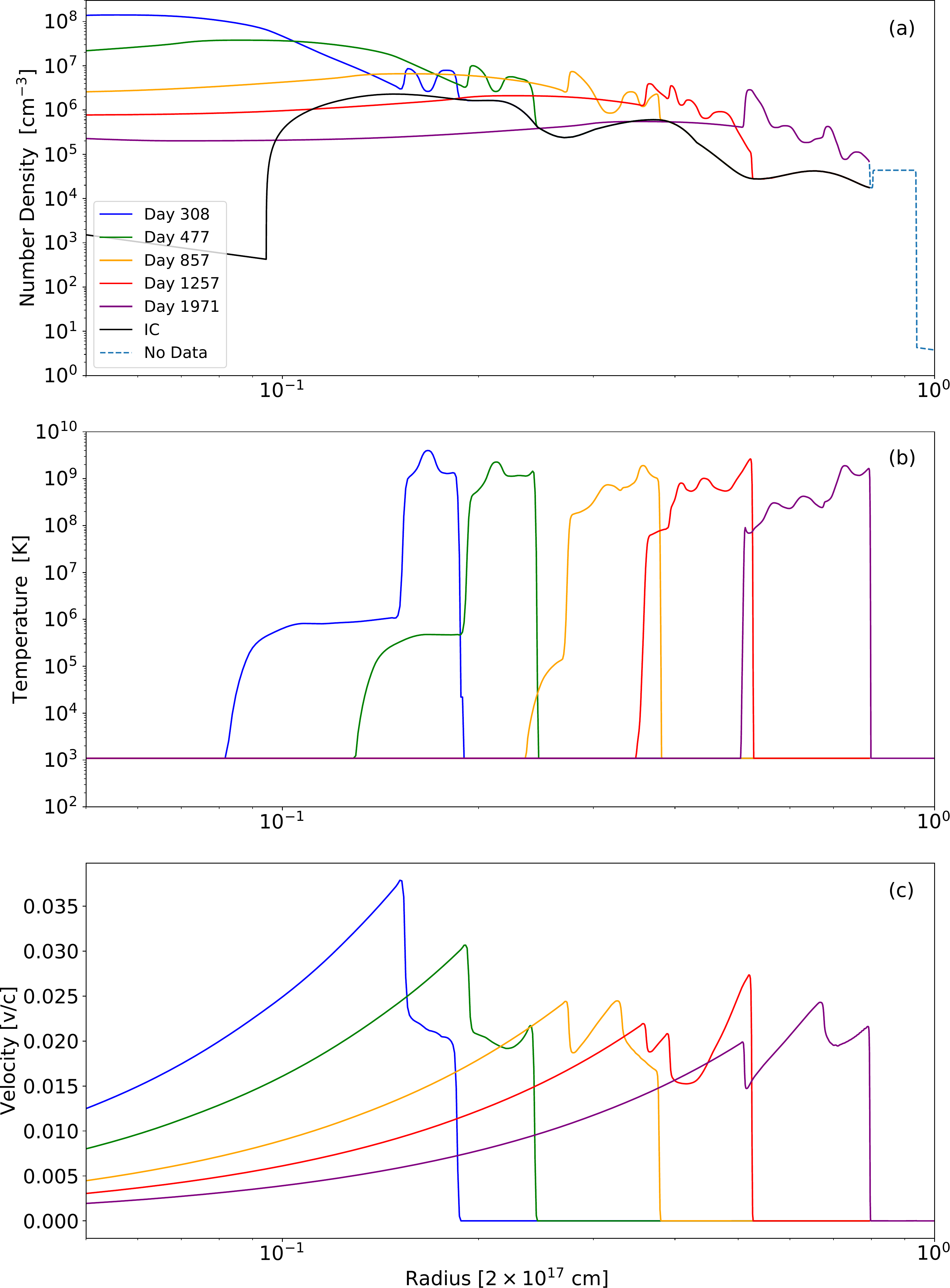}
\caption{The same as Figure \ref{fig3}, but for larger radii, showing in detail the interaction between the SN ejecta and the shell. The lines corresponds to epoch in which there are X-ray observations available. ``IC'' represents the initial shell density profile. In addition to nine densities corresponding to the epochs with X-ray data available, the amount of unshocked neutral mass is estimated by considering bound-free X-ray absorption. As the absorption does not depend on the density stratification but only on the amount of mass crossed by the X-rays, we show this region as uniform in the figure (dashed line in the top panel).
}
\label{fig5}
\end{figure}

The interaction of the ejecta with the shell is shown in Figures \ref{fig4} and  \ref{fig5}. The shell presents large density fluctuations (see the upper panel of figure \ref{fig5}). Then, the shock propagation leads to the formation of strong reverse shocks which interaction produces the complex shock structure observed in Figure \ref{fig5}. 
Once interacting with the shell, the shock velocity quickly drops to $\sim 7500$ km s$^{-1}$\footnote{From the Hydrogen line, a velocity of $\sim$ half of this value is inferred \citep{milisavljevic15}. This is consistent with the hydrogen lines being produced in the post shock clumpy medium or by recombination of the upstream medium.}, then maintain an approximately constant velocity (see Figure \ref{fig4}). Small fluctuations in the shock velocity are present at $\sim 10^3$ days (see the bottom panel of Figure 3), with increases (drops) in velocity by $\sim 3000$  km s$^{-1}$ corresponding to drops (increases) in the density. 

Initially, the reverse shock is stronger then the forward shock. Thus, the shocked ejecta is hotter than the shocked wind (Figure \ref{fig5}, blue and green lines in the middle panel). As the ejecta crosses the reverse shock and approaches the reverse shock velocity, the RS temperature drops becoming smaller than the FS temperature. Thus, the bremsstrahlung specific emission (larger at smaller temperatures) is initially dominated by the shocked wind, and later mostly originated into the shocked ejecta.

 A fit to the density profile gives $\rho \propto r^{-3.00\pm 0.06}$ (although there are large fluctuations), consistently with the constant shock velocity seen in Figure \ref{fig5} (as $E\sim M v^2\sim R^3 \rho v^2$ implies that the velocity is constant as long as $\rho\propto R^{-3}$ and the shock is adiabatic). The shell mass is 2.6 M$_\odot$. This value is consistent with the 3.0 M $\pm$  0.6 M$_\odot$ (VLB model) determined by \citet{brethauer21} which also assumes spherical symmetry and within the range of typical shell masses observed in type IIn SNe (0.1-10 M$_\odot$, see, e.g., \citealt{smith17, branch17}). 


To determine the structure of the shell, we left as a free parameter in the GA algorithm the amount of neutral hydrogen still to be crossed by the ejecta (inferred from the bound-free absorption). We get M $=0.38$ M$_\odot$ solar masses for this component at t = 1971 days, which is represented by a constant density dashed line in the top panel of Figure \ref{fig5}. We notice that the exact structure of this region can not be determined. Nevertheless, it will extend to $r\sim 2.3 \times 10^{17}$ cm if the shell density continues dropping as $r^{-3}$, in which case (moving at a constant speed as discussed above) the SN shock will break out of it $\sim 8.5$ years after the explosion. Late X-ray and radio observations will help to constrain the outer structure of the shell. 

\subsection{Radiation}
\label{Rad}

To compare the model with observations, we compute the X-ray bremsstrahlung emission coming from the shocked material. We assume that the shocked material is completely ionized\footnote{This assumption is justified by the large post-shock temperature and the presence of strong photo-ionizing X-ray and UV emission.} (so that it does not contribute to the bound-free opacity) and that the unshocked shell is neutral. Extending the radio synchrotron emission to X-rays by assuming $F_{\nu} \propto \nu^{-(p-1)/2} \propto \nu^{-1}$ with $p\sim 3$, \citet{margutti17} showed that the synchrotron emission gives a negligible contribution by X-ray flux (i.e., $\sim 2$ orders of magnitude smaller than the observed values), suggesting a thermal origin for the X-rays.

\begin{figure}[]
\centering
\includegraphics[width=0.9\linewidth]{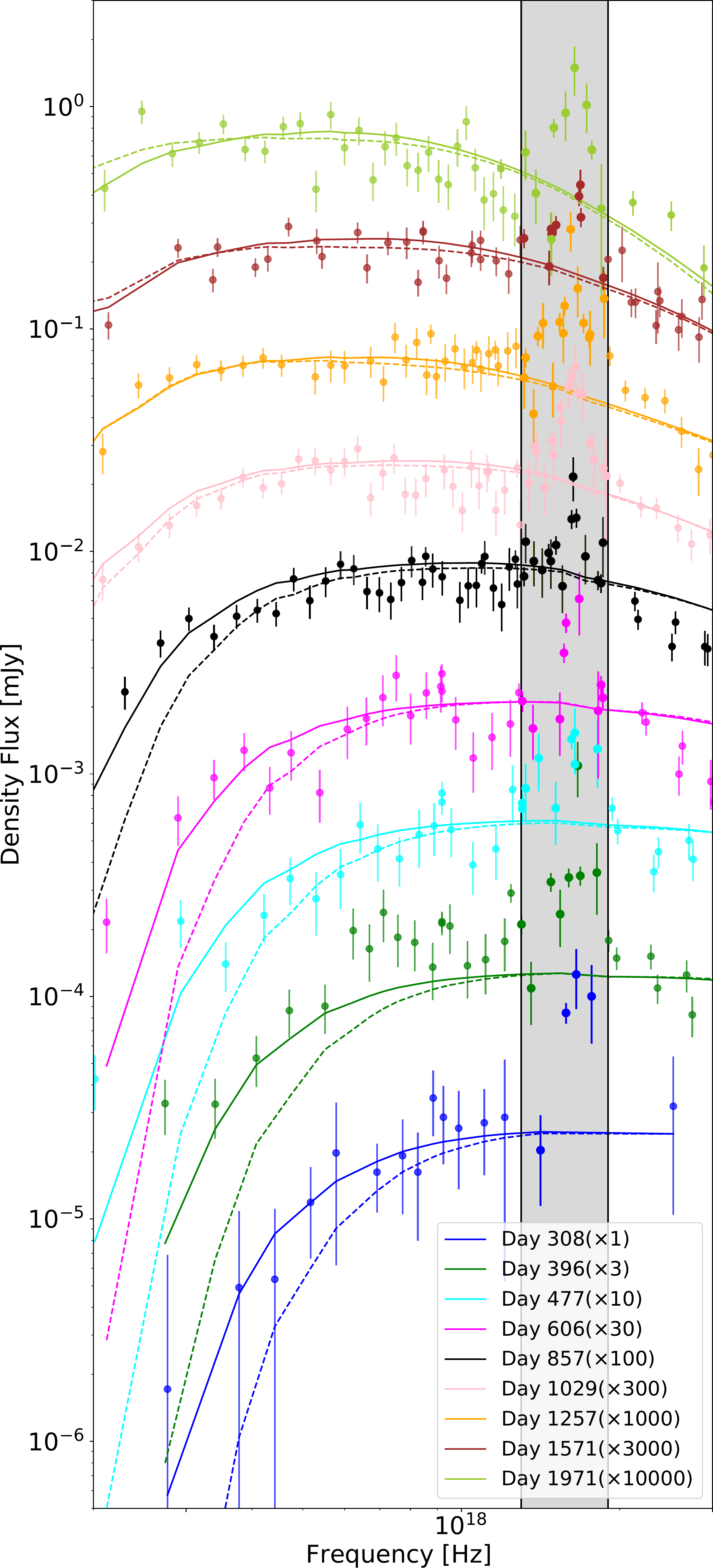}
\caption {Comparison between X-ray observations and synthetic observations computed for the best model. Full lines represents solar metallicity while dashed lines are for half solar metallicity. We do not include the shaded grey area in the fit, as it is dominated by Fe emission lines. Curves at different times are rescaled so they do not overlap with each other.}
\label{fig6}
\end{figure}

Figure \ref{fig6} shows the X-ray spectra at 308, 396, 477, 606, 857, 1029, 1257, 1571, 1971 days and the best fit obtained by employing the GA (Section \ref{GA}).
Dashed lines are computed by assuming solar metallicity in the neutral shell. 
The best fit, in this case, predicts a larger absorption by a factor  $\lesssim 5$ at $2\times 10^{17}$ Hz than observed.
Solid lines, which fits better the observational data, are computed by assuming that the neutral medium has half of solar metallicity.

\section{Discussion}

The optimization method employed in this work allows us to determine the detailed structure of the shell. 
As described in the previous section, the shell has a mass of 2.6 M$_\odot$ (2.2 M$_\odot$ of shocked and 0.4 M$_\odot$ of still unshocked gas), a density stratification $\rho \propto r^{-3}$ and extends from $1.6\times 10^{16}$ cm to  $1.87\times 10^{17}$ cm.
While uncommon in type Ib SNe, SN IIn show evidence of strong interaction, with shell masses of $0.1-10$ M$_\odot$ \citep[see, e.g.][and references therein]{smith17, branch17}. Also, many type IIn SNe show an X-ray emission inconsistent with a density profile $\rho\propto r^{-2}$, implying a steeper density stratification \citep{dwarkadas12}. While sharing many characteristics with SN IIn, in the case of the SN\,2014C the shell is located at a much larger distance, implying that it was ejected $\sim 60/(v_w/100 {\rm \; km \; s^{-1}})$ yrs before the SN 
explosion.

\citet{harris20} presented a series of numerical simulations of a SN ejecta interacting with a wall located at $r_0 \sim 1.6\times 10^{16}$ cm, and with a CSM wind at larger radii. They deduce a small mass for the wall ($\sim 0.04-0.31$ M$_\odot$), in apparent contradiction with previous estimations \citep[e.g.,][]{margutti17}. Actually, the density of the CSM wind is assumed to be a factor 4-7 smaller than the density of the wall at $r_0$. The CSM wind, then, corresponds to a mass loss $\dot{M}_w \sim 10^{-2}$ M$_\odot$ yr$^{-1}$ (assuming $v_w=10^8$ cm s$^{-1}$). Our results (which, we stress, have been obtained by a ``blind'' fit, i.e. without assumptions on the final shell structure) clarify this inconsistency, showing that the wall and the CSM wind are both part of the same extended structure (see figure \ref{fig5}), with the densest shell material located at $\sim 2\times 10^{16}$ (see Figure \ref{fig5}) in agreement with \citealt{harris20}, and the outer shell density dropping quickly with radius. Thus, we conclude that the same event is responsible for the ejection of the full massive shell.

The parameters determined for the SN\,2014C are remarkably similar to those inferred for the SN 2001em. The X-ray, radio and H$\alpha$ emission from SN 2001em have been interpreted as evidence of interaction with a 3 M$_\odot$ hydrogen-rich shell \citep{chugai06}. VLBI observations showed that the shell is located at $7\times 10^{16}$ cm, and expanding with a velocity of $5800 \pm 10^4$ km s$^{-1}$ \citep{bietenholz07}, which is consistent with the $7500$ km s$^{-1}$ inferred here.

The best fit to the data is achieved by considering a bound-free cross-section, corresponding to half solar metallicity.
This low metallicity is in contradiction with observations of the Fe line. The prominent Fe emission line at 6.7-6.9 keV is consistent with a metallicity larger by a factor of $\sim 5$ with respect to solar metallicity. This can be reconciled with our results by assuming that the medium is clumpy, with high density (and lower temperature) regions responsible for most of the Fe emission \citep{margutti17}.

An equivalent effect is obtained considering that the X-ray emission and/or UV radiation coming from the shocked shell and SN ejecta can partially ionize the unshocked material, producing an effect similar to a drop in the metallicity, as the optical depth is $\propto n_{H^0} \sigma_\nu$, being $n_{H^0}$ the density of neutral hydrogen and $\sigma_\nu$ the bound-free cross-section (which depends on metallicity). In this case, we expect the mass of the shell to be larger than the value obtained in this paper, although by a small factor as the bremsstrahlung emission is $\propto n_e^2$. A detailed calculation of the ionization of the shell is left for a future work.


Different origins for the shell have been considered. The possibility of the shell being due to a massive wind ejection \citep[e.g.,][]{deJager88, Leitherer09, kuriyama20} is unlikely, as it would correspond to an extremely large mass loss rate of $\sim 10^{-2}$ M$_\odot$ yr$^{-1}$
($v_w$/100 km s$^{-1}$). Other possibilities include a sudden outburst some time before collapse, which would remove the most external layer of the star where almost all hydrogen is found \citep{Smith13b}, or binary system interactions in which the envelope of the most massive star has been stripped away \citep{Sun20}. A better understanding of the origin of this ejections can be achieved only by detailed theoretical models coupled to a larger sample of observed interacting supernovae.

The density fluctuations have a periodicity of $\sim 4\times 10^{16}$ cm. A similar periodicity has been observed in the radio emission from SN\,1979C \citep{weiler91} and have been interpreted as evidence of a binary system in which the orbital motion modulates the wind density \citep[see][]{yalinewich19} which interacts with the stellar outburst. If this is also the case of SN\,2014C, it would imply that the binary system is very detached (as the binary period is $\gg 4\times 10^{16}/v_w \sim 10$ yrs $(v_w/10^{8})^{-1}$). The companion star would then be not responsible for the loss of the envelope of the primary star. An alternative explanation is that the density fluctuations seen in the GA fit are the direct result of a modulation in the outburst from the progenitor star.

Finally, we notice that the simulations presented here assume that the shell is spherically symmetric. This is consistent with VLBI observations \citep{bietenholz18}. Small scales inhomogeneities are expected. The density profile shown in Figure \ref{fig5} show large scale density fluctuations. Furthermore, it is likely that the medium is, at some scale, clumpy. If the shell is not perfectly homogeneous before interacting with the SN ejecta, the interaction will amplify the inhomogeneities, leading to a multi-phase medium with denser/colder regions in pressure equilibrium with more tenuous/hotter regions, which is consistent with the strong Fe emission line observed at $\sim$ 6.5 keV  \citep{margutti17}. Also if the shell is initially nearly perfectly homogeneous, the contact discontinuity which separates the shocked SN ejecta from the shocked shell is prone to Rayleigh-Taylor (RT) instabilities, so that we can expect the formation of plasma filaments and dishomogeneities in the post-shock region which can not be captured by our numerical simulations (but see \citealt{harris20} for an approximated treatment of RT instabilities in one dimensional simulations). As the bremsstrahlung emission is $\propto Z^2 n^2$, inhomogeneities in the shell and mixing with the higher metallicity ejecta lead to a larger emissivity, implying that the mass of the shell should be taken as an upper limit.

\section{Conclusions}

In this paper, we presented hydrodynamical simulations of the strongly interacting SN\,2014C. First, we follow the propagation of a SN shock through the progenitor star. Then, by using as input the outcome of the small scale simulation (i.e., density, pressure and velocity profiles), we run a large set of simulations. As described in section \ref{GA}, we initialize the shell with a uniform density $n_{\rm shell} =10^7$ cm$^{-3}$. We follow the propagation of the SN shock as it interacts with the wind launched by the progenitor Wolf-Rayet star and with the massive shell.
We compute the bremsstrahlung emission using the algorithm described in Section \ref{free}, and compare the results with observations. At each step, we run a large number of simulations changing the shell density profile.
As a result, we determine the shell structure and metallicity. In particular, we get a mass of $2.6$ M$_\odot$ for the shell and a density profile $\rho\propto r^{-3}$. We also found that the shell is very extended, with a size $\gtrsim 10^{17}$ cm. If the shell stratification continues with the same slope, the SN shock will break out of it nearly 8 yrs after the explosion, i.e. during 2022.

Radio and X-ray emission allows us to understand the mass loss history of core-collapse SNe progenitor on time-scales which are impossible to study by direct observations. 
As we have shown in this paper, optimization methods can be used, coupled with hydrodynamical simulations, to model the density stratification of the environment once data at several epochs are available, as in the case of SN\,2014C. The X-ray emission tracks the forward and reverse shock emission, depending on the density of the environment and the ejecta velocity. The H$\alpha$ emission tracks the shocked shell and the unshocked medium fotoionized by the X-ray and UV radiation. All together, a detailed fit of the different components can help us to get a better understanding of this system.
Then, coupled with detailed modeling of the radio emission, this analysis can allow us to determine the microphysical parameters as a function of time (which are usually degenerate with the density of the environment and ejecta velocity), giving us direct information on the particle acceleration process. In this paper, we describe this technique by analyzing the X-ray bremsstrahlung emission. The extension to radio and optical emission will be considered in a future study.

\acknowledgements
We thank  Luc Binette, Cesar Fern\'andez Ram\'irez, Leonardo Ferreira and Claudio Toledo Roy for useful discussions. FV and FDC acknowledge support from the UNAM-PAPIIT grant AG100820.
We acknowledge the computing time granted by DGTIC UNAM (project LANCAD-UNAM-DGTIC-281).
R.M.~acknowledges support by the National Science Foundation under Award No. AST-1909796 and AST-1944985. Support for this work was provided by the National Aeronautics and Space Administration through Chandra Award Number GO9-20060A  issued by the Chandra X-ray Center, which is operated by the Smithsonian Astrophysical Observatory for and on behalf of the National Aeronautics Space Administration under contract NAS8-03060.



\end{document}